# Reconstructing a B-cell clonal lineage. I. Statistical inference of unobserved ancestors

Thomas B. Kepler
Department of Microbiology, Boston University School of Medicine, 72 E Concord St, R504D, Boston MA 02118 & The Hariri Institute for Computing and Computational Science & Engineering, Boston MA.
tbkepler@bu.edu

## Abstract

One of the key phenomena in the adaptive immune response to infection and immunization is affinity maturation, during which antibody genes are mutated and selected, typically resulting in a substantial increase in binding affinity to the eliciting antigen. Advances in technology on several fronts have made it possible to clone large numbers of heavy-chain light-chain pairs from individual B cells and thereby identify whole sets of clonally related antibodies. These collections could provide the information necessary to reconstruct their own history—the sequence of changes introduced into the lineage during the development of the clone—and to study affinity maturation in detail. But the success of such a program depends entirely on accurately inferring the founding ancestor and the other unobserved intermediates.

I have developed an Empirical Bayes method that allows one to compute the posterior distribution over ancestors, thereby giving a thorough accounting of the uncertainty inherent in the reconstruction. I demonstrate the application of this method on heavy-chain and light-chain clones, assess the reliability of the inference, and discuss the sources of uncertainty.

## Background

During the course of an infection, the host's immune system produces antibody molecules that bind to molecular determinants (antigens) on the infectious agent, thereby neutralizing the agent and targeting it for removal by additional antimicrobial effectors. The heavy and light chain immunoglobulin (Ig) genes that encode the components of the antibody molecule result initially from the stochastic intrachromosomal rearrangement of gene segments arrayed in libraries of such gene segments [1]. These genes are further modified after the activation of the B cells that possess them through somatic hypermutation targeted to the rearranged Ig genes [2]. Those B cells whose Ig genes encode molecules with greater affinity for the eliciting antigen gain a proliferative and survival advantage. In this way, the overall affinity of the pool of serum antibodies increases, sometimes by two or more orders of magnitude. This *affinity maturation* [3] is an essential component of the establishment of humoral immunity, the basis for the large majority of successful vaccines [4].

A great deal has been learned about affinity maturation, particularly with regard to the mechanism of somatic hypermutation [5] and the dynamic organization of the cellular environment in which affinity maturation takes place [6,7] (for a recent review, see [8]), but the mechanism underlying the selective aspects of affinity maturation remains poorly understood. There is increasing interest in the manipulation of affinity maturation pathways in vaccinology [9] and thus in comparing the biophysical properties of mature antibodies to those of their inferred unmutated ancestors (UA) [10-18]. Little attention has been paid, however, to the uncertainties inherent in the inference of these UAs. Given the sensitive dependence of antibody-antigen interactions on single amino acid changes [19], estimating these uncertainties is essential. Under some circumstances, there may be more than one history consistent with prior knowledge that is supported by the data; having the means to determine these cases and provide a set of alternative UAs that as an ensemble cover a significant posterior probability could be valuable, as was shown by Alam et al. in a study of the affinity maturation of a broadly neutralizing anti-HIV-1 antibody [14].

The inference of ancestral rearrangements involves the alignment of two (light chain) or three (heavy chain) gene segments in tandem to the target mature Ig gene. The identities of the gene segments are not known in advance. Instead, there is a library of gene segments from which each segment is drawn stochastically; the identity of each segment is part of the inference. The problem is complicated by randomness in the location of the recombination points, where each gene segment begins or ends, because this condition implies that the alignments are not independent. Further challenges are encountered by the presence of nontemplated (N-) nucleotides added at random to the junctions between gene segments, and of course, by point mutations.

There is a well-developed literature on ancestor reconstruction in phylogenetics (see, e.g. [20]). These methods have informed the development of our methods, but are not themselves sufficient to the problem at hand. The difference between these phylogenetic methods and the method described here is that the previous methods do not take into account the complex process through which the Ig ancestor is constructed. This process places a strong statistical constraint on what ancestral states are permissible. My method owes a great deal to this prior work but does not aim to improve upon it fundamentally. It simply extends a small part of its methods to a new domain of application.

Independent of this previous work from phylogenetics there are applied methods developed by computational immunologists. Indeed, computational methods developed to address the problem have been used for some time [21]. There are several different approaches and corresponding programs available online for carrying out these analyses, including iHMMune [22], V-Quest [23], Joinsolver [24], SoDA [25] and SoDA2 [26]. None of these applications, however, provides either of two features essential for the systematic reconstruction of clonal histories. First, one must be able to use all of the information available in a set of clonally related Ig genes in a statistically principled manner. All currently available Ig alignment tools work with one sequence at a time. Second, one needs systematic uncertainty estimates on the UA. In order to say anything of interest about the UA and the clonal history, there must be some level of certainty that the inferred sequence really is the actual UA.

The method described here provides these features. It is based on a hierarchical model of Ig gene development that produces an analysis of the clonal history and posterior probabilities on the UA. The method uses the information available across all members of a clone in a consistent and powerful manner.

## Methods

We start with a query set $Q$ of observed Ig variable-region gene sequences assumed to share descent from a common ancestor $\alpha$. The task is to estimate the DNA sequence $\alpha$ or, more generally, a posterior probability on $\alpha$. There are two distinct stochastic processes that together give rise to *Q*. The stochastic intrachromosomal rearrangement process transforms the germline configuration to the unmutated (naïve) ancestor. Somatic mutation transforms the naïve ancestor to the mature (mutated) antibodies that are observed. To each of these stochastic processes there corresponds a probability function, each of which, in turn, has a natural interpretation within the framework of Bayesian inference. The rearrangement process generates a distribution $P_0(\alpha)$ on unmutated ancestors. For each unmutated ancestor $\alpha$, somatic mutation then generates the likelihood function $P(Q \mid \alpha)$ relating the ancestor to the observed query sequences. Once these functions are computed, Bayes' Theorem is used to compute the posterior probability on $\alpha$ given *Q*,

$$P(\alpha \mid Q) = \frac{P(Q \mid \alpha) P_0(\alpha)}{\sum_{\alpha'} P(Q \mid \alpha') P_0(\alpha')} \qquad (1)$$

*Parameterization of the recombination process*

To avoid unnecessary complication, we will use light chain sequences for illustration. The extension to heavy chains is straightforward, but even for the simpler light chains the notation becomes clumsy and obscures the intuition behind the method. Heavy chain rearrangements involve an additional gene segment (DH) and two junctions rather than the one that light chain rearrangements have. Figure 1 illustrates the parameterization of a heavy-chain rearrangement and provides a guide applicable to both heavy and light-chain rearrangements.

A light-chain rearrangement results from the selection of a V-gene segment *V*, the selection of a J-gene segment *J*, the specification of the recombination point in both of these segments $R_V$, $R_J$, and the sequence *n* of the N-nucleotides randomly added to the junction between the gene segments. We regard these elements as parameters in a statistical model: *V* and *J* are categorical parameters naming specific gene segments, $R_V$ and $R_J$ are integers, and *n* is a DNA sequence. $R_V$ is defined as the position of the $3'$-most V nucleotide included in the rearrangement; $R_J$ is the position of the $5'$-most J nucleotide included. The DNA sequence *n* may have length zero (meaning that the V and J segments are directly joined and no N-nucleotides occur).

Each combination of parameter values generates a specific DNA sequence, although a given sequence may be generated by more than one set of parameter values. Our strategy is to compute the posterior distribution on these parameters, and use it to generate posteriors on the quantities of interest, such as the nucleotides in the founder.

Let $S(V, J, R_V, R_J, n)$ be the sequence generated by indicated arguments. Then the distribution on unmutated rearrangements is

$$P_0(\alpha) = \sum_{V,J,R_V,R_J,n} I\left[\alpha = S(V, J, R_V, R_J, n)\right] \pi(V, J, R_V, R_J, n) \qquad (2)$$

where $I$ is the Boolean indicator: $I[true]=1$, $I[false]=0$ and $\pi$ is the prior probability on rearrangement parameters.

We take $V$ and $J$ to be independent and $\pi(V,J,R_V,R_J,n)=\pi(V,R_V)\pi(J,R_J)\pi(n)$. Although this assumption is not strictly true—there are small correlations among V, D, and J gene segments [27], the inclusion of these correlation would have very small effects on the resulting inference at the cost of substantial computational effort.

For the analyses in this paper we use gene-segment libraries derived from the IMGT reference libraries [28]. These libraries contain multiple alleles for each gene segment locus. We assign priors to the gene segments such that each gene segment locus has the same prior probability, regardless of the number of allelic variants present. Within a gene-segment locus, the distribution on alleles is uniform. When more prior information is available—for example, if one knows the allelic frequencies in the relevant population or knows precisely which alleles are carried by the subject—this information is easily accommodated in the prior probabilities.

The recombination sites are also assigned prior probabilities uniformly across their assumed range. The largest allowed value for $R_V$ corresponds to the position just $3'$ of the codon encoding the second invariant cysteine residue. The largest allowed value of $R_J$ corresponds to the position just $5'$ of the codon encoding the invariant tryptophan residue. For all gene segments, the smallest allowed value of the recombination points is -4, corresponding to four P nucleotides [29].

For N-nucleotide sequences, we use an improper prior, formally assigning a uniform distribution across all sequence lengths. While this assumption, when taken as a statement about reality is clearly wrong, its consequences for inference are minor. is, ancestral sequences that have unreasonably long N regions will be judged very unlikely to give rise to the observed sequences and will not contribute substantially to inferences. The mechanics of this phenomenon will become clearer when we describe the computation of the likelihood and sequence alignment.

*The Likelihood Function*

The second probability function we require is the likelihood, describing the probability that the query sequences Q arose from a given ancestor $\alpha$ by somatic mutation. The likelihood function depends implicitly on the multiple sequence alignment used as well as on the phylogenetic tree. It is computationally infeasible to account completely for these additional sources of uncertainty. Indeed, it remains a significant challenge in the general case [30]. Fortunately, somatic hypermutation only infrequently creates insertions or deletions [31], which are the major cause of uncertainty in multiple sequence alignment. With regard to uncertainty in the phylogenetic tree, it has been shown that inference of ancestral states is relatively insensitive to variation in the assumed tree [32].

For the alignment, we assume that the complete multiple alignment $A_C$ can be decomposed into a multiple sequence alignment $A_Q$ among the query sequences in Q and the alignment $A$ between $A_Q$ and the UA. We estimate $A_Q$ in advance and treat it as given in the subsequent computations. Then for each gene segment, we compute the maximum likelihood alignment between it and $A_Q$.

Each tree $T$ is represented by a tree $T_1$ with average branch length 1 and a mutation rate μ taken to multiply each branch of $T_1$ to yield T. Although the estimated ancestor is insensitive to variation in the assumed tree [32], our own observations show that the estimate of uncertainty is sensitive to the assumed overall mutation rate, i.e., to the overall scaling of the branch lengths.

Our procedure is to iteratively estimate $T_1$ given the UCA and the UCA given $T_1$, integrating over μ at each stage. We start with a simple $T_1$ invariant under permutations of the gene assignments to tips (a *palm tree*), Then, given $T_1$ estimate the posterior on the rearrangement parameters (integrating over μ). Find the UCA with maximum posterior likelihood, and use this sequence at the root to re-estimate $T_1$. Continue iteratively until convergence is reached.

Although the pairwise alignments $A_V$, $A_D$, and $A_J$ of the V, D and J gene segments to Q are not independent, they are conditionally independent given the recombination points. Therefore, the likelihood factorizes into components corresponding to gene segments as follows, using the light chain for the example,

$$\begin{aligned} &P(Q|V,J,R_V,R_J,n,A,T)\pi(V,J,R_V,R_J,n)= \\ &\quad P(Q|V,R_V,A_V,T)\pi(V,R_V)P(Q|J,R_J,A_J,T)\pi(J,R_J) \\ &\quad \times P(Q|n,T)\pi(n)f(R_V,R_J,A_V,A_J) \end{aligned} \quad . \qquad (3)$$

The last function contains the dependence among the gene segment pairwise alignments. $f(R_V,R_J,A_V,A_J)=1$ when the position of $R_J$ in $A_Q$ is 3' of the position of $R_V$ in $A_Q$, that is, when the gene segments do not overlap. Otherwise, it is zero.

*Sequence alignment and somatic mutation*

We take the positions in the ancestor to evolve independently. For a single query sequence $q$, we have

$$\log P(q|\alpha,\lambda)=\sum_{i=1}^{L}\log M(q_i|\alpha_i,\lambda) \qquad (4)$$

where $q_i$ is the nucleotide at position $i$ in the query, $L$ is the length of $q$, $\alpha_i$ is the nucleotide at position $i$ in the ancestor, and $\lambda$ is the product of time and mutation rate, or branch length. The function $M$ represents the substitution model. For this paper, we will use the simple Jukes-Cantor form [33].

Within each component of the likelihood, the substitution model allows us to compute the likelihood for any sequence α placed at the root of $T$, conditional on $T$. Since the columns of the individual gene segment alignments are independent, the overall likelihood is the product of the likelihoods for each column in the alignment, each of which is given by taking the product of the likelihoods along each branch in $T$ and summing over all combinations of nucleotides at the interior nodes [34].

Given a pair of nucleotide sequences with one taken to be derived from the other, an alignment between them is equivalent to an accounting of the mutations via which the derivation occurred. Given a substitution model, there is an alignment scoring scheme that corresponds to that substitution model, so that the score for any alignment is the log of the likelihood of the corresponding set of substitutions.

The generalization of these observations to the alignment of a nucleotide sequence against a set of sequences pre-aligned among themselves is straightforward. Let the set of nucleotides at position *i* in the alignment be denoted $\mathbf{q}_i$, and the nucleotide in the ancestor at position *i* be denoted $\alpha_i$. We have the following pairwise alignment scoring scheme.

**Match score**—aligning the *j*th position in the ancestor against the *i*th position in the derived gene:

$$m(i,j) = \log M\left(\mathbf{q}_i \mid \alpha_j, T\right) \qquad (5)$$

**Insertion score**—aligning a gap in the ancestor against the *i*th position of the derived sequence:

$$I(i) = \log M\left(\mathbf{q}_i \mid -, T\right). \qquad (6)$$

**Deletion score**—placing a gap at any position in the derived sequence:

$$d(x) = \log M\left(- \mid x, T\right), \qquad (7)$$

where *x* is any nucleotide. To account for long deletions or insertions one could use an affine gap score, but in this paper just the simple gap penalties above are used.

*Nontemplated Nucleotides*

In addition to the standard scoring elements for pairwise alignment, the alignment of rearranging antigen receptors requires an additional scoring element for the treatment of N nucleotides. We will compute a score for the assignment of a given nucleotide to a generic N nucleotide rather than to a specific N nucleotide state (A,G,T,C). Denoting by $\pi_N(x)$ the prior probability for a random N-nucleotide to have state *x*, the score for asserting that the position *i* in the derived sequence is encoded by an N-nucleotide is

$$N_i = \log \sum_{x \in \{A,G,T,C\}} M\left(\mathbf{q}_i \mid x, t\right) \pi_N(x) \qquad (8)$$

For the analyses conducted in this paper, we take $\pi_N(x) = 1/4$ for all nucleotides *x*, though, again, the use of informative priors is straightforward.

With all the components of the scoring function in place, we are able to use dynamic programming to find the alignment that maximizes the alignment score.

*Algorithm*
The algorithm is schematized as follows.

(Preparation)
Align Q using multiple sequence alignment to give $A_Q$.
Assume an initial unit-length palm tree, $T_1$.

While not converged:
**{**

Estimate rearrangement parameters given $T_1$.
For each discretized value of μ
{
Compute the likelihood for each $\alpha_i \in \{A, C, G, T\}$ at each position $i$ of $A_Q$.
Align each gene segment $V$, ($D$), $J$ in the gene segment library to $A_Q$, using Eqs.(5-8), computing the likelihood for the relevant parameters in each alignment.
Compute the posterior on α conditional on μ using Eqs.(1,2).
}
Compute the posterior on μ.
Marginalize the posterior on α over μ.
Add the modal (maximum posterior probability) UA $\alpha^*$ to Q.
Estimate new tree $T_1'$ with $\alpha^*$ at root.
If $T_1' == T_1$ converged = true
Else $T_1 = T_1'$

}

Because of N-nucleotides and increased uncertainty estimating DH gene segments, CDR3 is typically the region of lowest confidence. In addition, the CDRs are the locations that accumulate mutations most rapidly in both selected and unselected genes [27]. For these reasons, CDR3 is susceptible to having its true mutation rate underestimated. We therefore use a mutation frequency 2-fold higher in CDR3 than in the remainder of the gene. This value is consistent with the enhancement of mutation frequency measured in CDR1 and CDR2 where there is much greater confidence in the counting of mutations [35].

The foregoing method was implemented using CLUSTALW [36] to compute $A_Q$, PHYLIP's dnaml [37] for clonal tree estimation, and our own software for all other computations.

## Results

To examine the reliability of error estimation for the method, we identified two relatively large sets of clonally-related genes for testing. The first, Clone H, is a set of 84 heavy-chain genes [38] of common length 376 nucleotides (nt), with an average (± standard deviation) pairwise difference of 30.4 ± 9.4 nt and a maximum pairwise distance of 61 nt. Fig. 2 shows the clonal phylogram for this set of sequences. The second, Clone K, is a set of 12 kappa-chain sequences [16] of length 299, with an average of 12.2 ± 4.8 nt differences and a maximum pairwise distance of 21 nt.

We applied the inference procedure to Clone H and found that the VH gene segments with the greatest posterior probabilities are VH4-34*01 and VH4-34*03, with nearly identical posterior probabilities of 0.49 each. These two alleles differ from each other in two places. The majority of sequences in the alignment matches one of the alleles at one of these two informative sites but matches the other allele at the other informative site. The modal DH gene segment is DH6-6*01 with posterior probability 0.94. The modal JH gene segment is JH6-1*02 with posterior probability greater than 0.99. The most likely rearrangement has VH using as many as 7 p-nucleotides, no VD n-nucleotides, and 14 DJ N nucleotides (Fig. 3). The observed sequences have an average mutation frequency of 8.0% compared to the UA.

The UA of Clone K is inferred to have been rearranged using VK1-39*1 with probability greater than 0.999 and to the JK1*1 with probability 0.98. No n-nucleotides are required for the rearrangement. The observed sequences have an average mutation frequency of 5.6% compared to UA.

The inference procedure produces a posterior marginal probability mass function over nucleotides at each position of the UA. The probable error at each position is defined as one minus the maximum value of the posterior probability at that position. The total probable error is the sum of the probable errors over positions, and gives the expected number of mismatches between the inferred modal UA and the true UA.

To examine the reliability of the estimated probable error, we subsampled the sequence sets and performed the inference on each of the subsamples. For Clone H, we generated ten pseudorandom samples for each size 1, 3, 9, and 27. For Clone K, we simply estimated the modal UA using each of the individual sequences alone. The resulting modal UAs were compared to the modal UAs inferred from the complete set.

For Clone H, the total probable error for the UA inferred from the complete set is 2.0. Figure 4 shows the results of these analyses for Clone H. The observed number of mismatches for each subsample is plotted against the total probable error for that subset. The distribution of probable error by nucleotide position shows that some uncertainty is attributable to uncertainty in the allele used in the ancestral rearrangement (Fig. 3, position 273) and some is attributable to uncertainty in the N nucleotides and junctions (Fig. 3, HCDR3).

For Clone K, the total probable error for UA inferred from the whole set is 0.07. For the 12 UAs obtained from individual sequences, the mean total probable error is 0.14 ± 0.05. There were no mismatches among the light-chain UAs.

*Influence of prior distribution*

To quantify the impact of the prior distributions on the inference, we performed the inference using the same sequence sets, but with a simple uniform prior on nucleotides at each position rather than the prior based on knowledge of the rearrangement mechanism and gene segments. Under this model, the modal UA differs from that of the full rearrangement-based model in 11 positions for the heavy-chain clone, and in 10 positions for the light-chain clone. The total probable error for the heavy chains and light chains is 8.5 and 11.5, respectively for the model with uniform priors.

## Discussion

We have developed a method for the inference of clonal history in sets of affinity-matured clonally-related immunoglobulin genes. The method allows one to compute posterior distributions on the rearrangement parameters, and hence marginal distributions on several elements, including the nucleotide sequence of the unmutated ancestor.

The probable error is strongly dependent on the interplay of N nucleotides and mutation frequency. This phenomenon occurs because nucleotides near the recombination junction are ambiguous with regard to their origin. A nucleotide that does not match the relevant gene segment at a position near the unknown recombination junction may have been encoded by the gene segment and mutated. Alternatively, it may have been encoded by an N nucleotide. The relative probabilities of these alternatives depend on the mutation frequency. If there are few mutations elsewhere in the gene (where they can be determined more reliably) the likelihood of a mismatch in the junction being due to a mutation is small.

The second major source of uncertainty is allelic diversity. It is often the case, as it is with Clone H, that mutation has destroyed the information required to distinguish which of two or more alleles was used. The greater part of the total uncertainty will be due to one of these two phenomena (Figure 4). This state of affairs also implies that the errors may be correlated, and the distribution of the total number of mismatches overdispersed, as is evident in Figure 5.

We expect the total uncertainty to be proportional to the distance from the root to the most recent common ancestor of the observed sequences. Adding related sequences to a clonal set improves the inference to the extent that they push back the time of the most recent ancestor.

Where there are few N-nucleotides and allelic polymorphism either not present or not obscured by mutations, the UA can be inferred with great precision, even in the presence of significant levels of mutation, as is the case with Clone K.

## Conclusions

Technology now provides immunologists with the means to reconstruct clonal histories, synthesize the unobserved ancestors, and retrace the steps of affinity maturation to provide deeper insight into the humoral immune response in general and into vaccine design in particular. But the value of the information obtained in this way is wholly dependent on the reliability of the inferential part of the reconstruction. If the ancestors and intermediates are misinferred, the reconstructed history will be potentially misleading.

The methods outlined here are intended to ensure reliable inference and to indicate when multiple histories must be considered.

## Acknowledgements

I thank Grace Kepler, Barton Haynes, Larry Liao and the members of the Duke Human Vaccine Institute Antibodyome group for stimulating discussions.
*Funding*: This work was supported by NIH/NIAID research contract HHSN272201000053C and a Vaccine Development Center grant in the Collaboration for AIDS Vaccine Discovery Program from the Bill and Melinda Gates Foundation (B. Haynes, PI).

## Figures

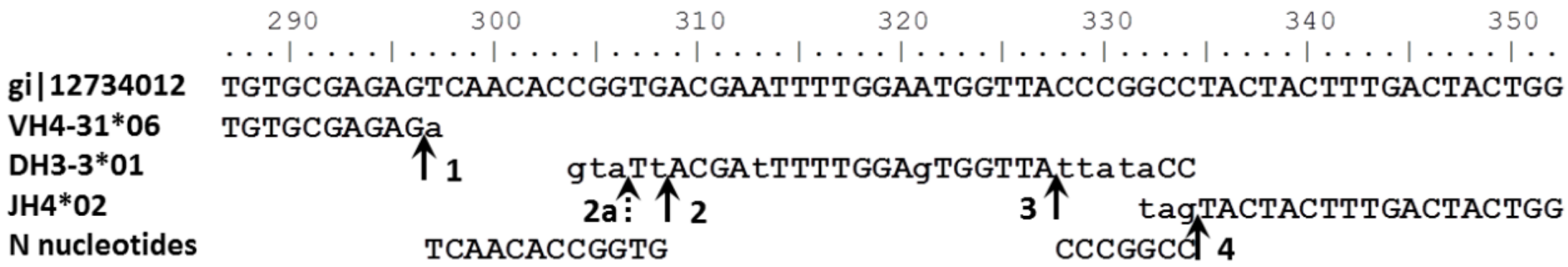

**Figure 1. Illustration of parameters for the rearrangement model.**

Labelled vertical arrows indicate the positions of the recombination sites: 1) RV = 1; 2) RD1 = 5, 3) RD2 = 7; 4) RJ = 3. The dashed arrow 2a indicates a possible alternative recombination site: RD1 = 3. Lower-case letters in the gene-segment sequences indicate mismatches between the observed sequence and the gene segment. The last line shows N nucleotide sequences consistent with the observed sequence.

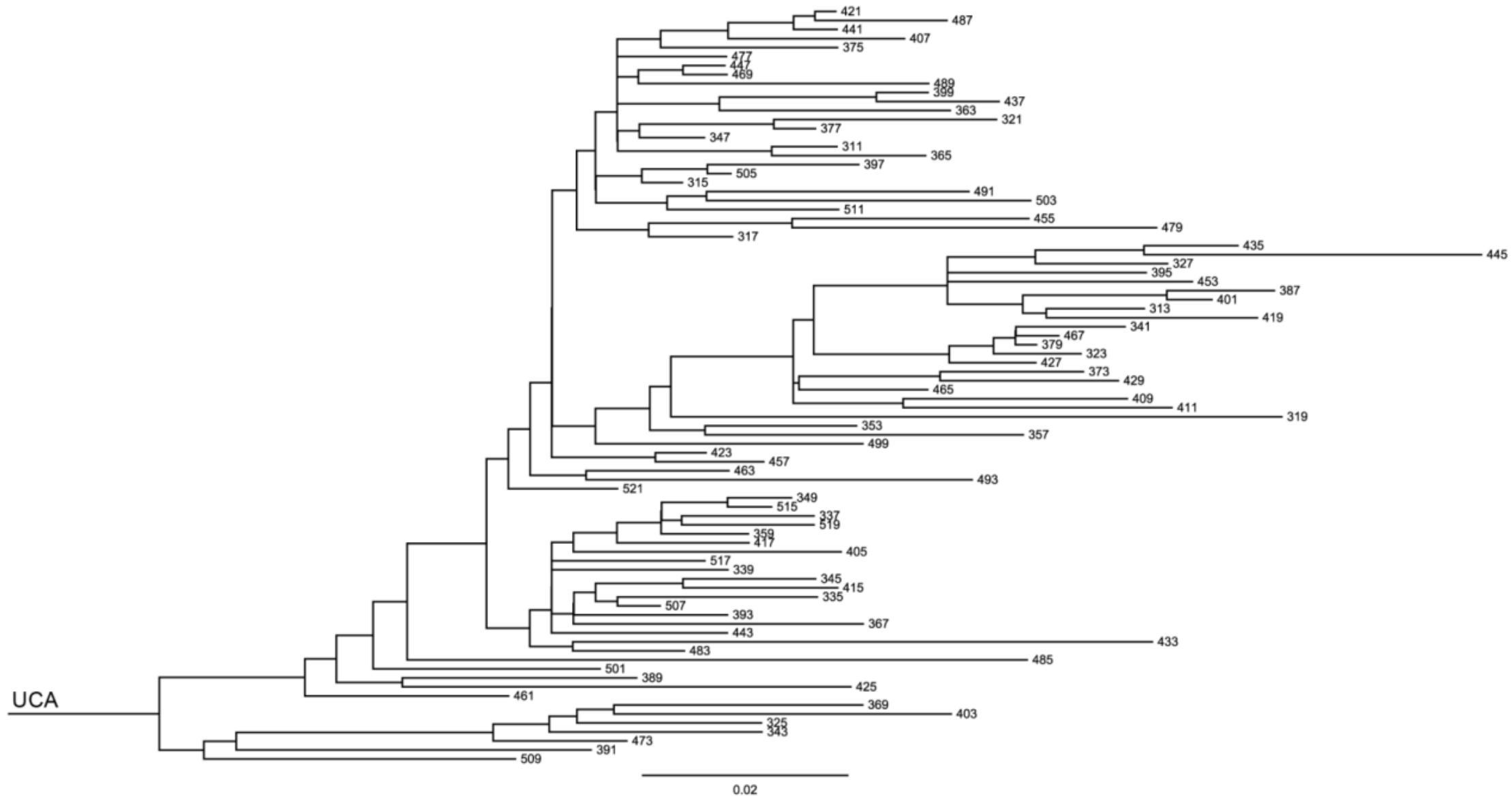

**Figure 2. Phylogram of Clone H.**
The scale bar shows evolutionary distance, or expected number of mutations per position.

**Figure 4. (NEXT PAGE) Nucleotide alignment and error profile.**
Nucleotide alignment of observed heavy-chain sequences, inferred unmutated ancestor, and modal gene segments, with the probable error (below), illustrating the influence of N nucleo-tides, junctions, and allelic ambiguity on uncertainty. The large probable error at position 273 is due to allelic ambiguity. A second position in FR1 has similar probable error due to allelic ambiguity (not shown). HCDR3 is indicated. The 84 sequences at the top of the alignment are fragments of the observed members of Clone H (naming is arbitrary). The 4 sequences at the bottom of the alignment are the modal UA, and the modal gene segments. A dot in the sequence indicates a match to the UA at that position.

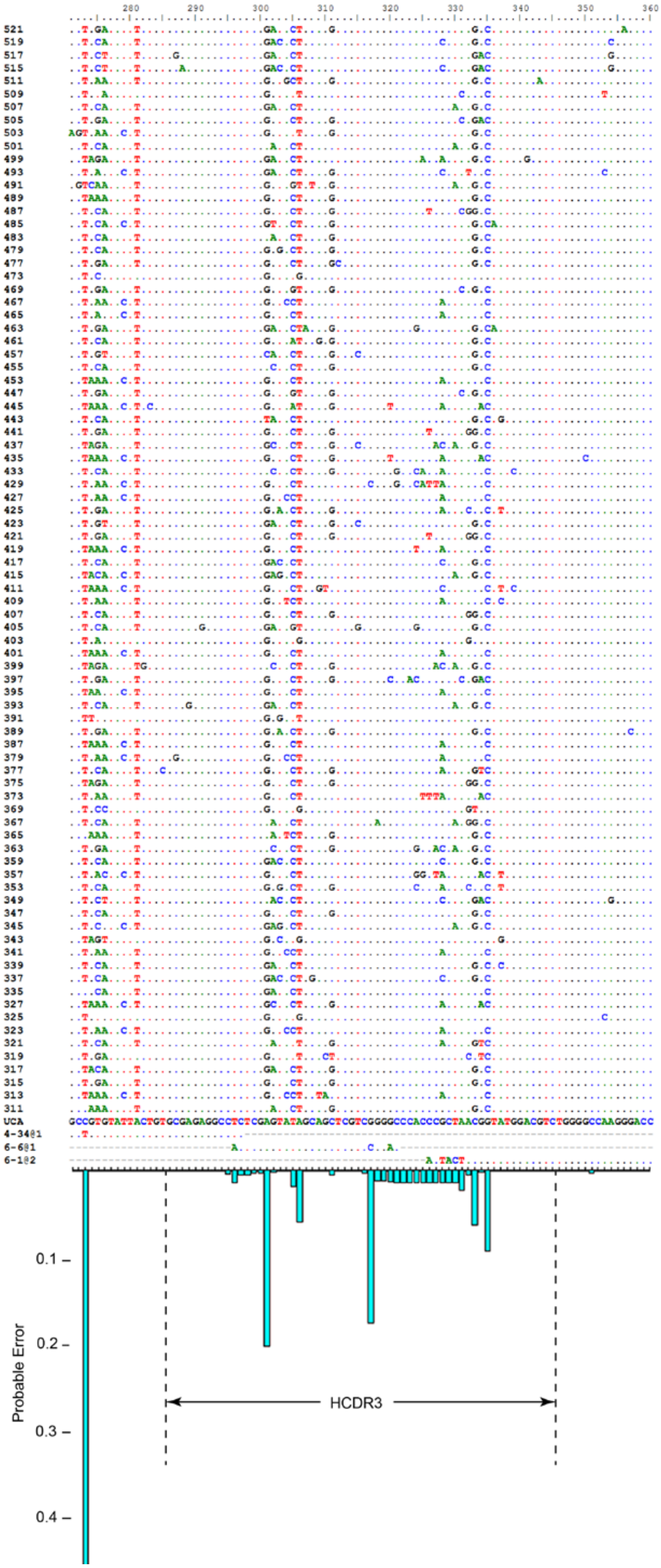

HCDR3
Probable Error
0.1
0.2
0.3
0.4
UCA

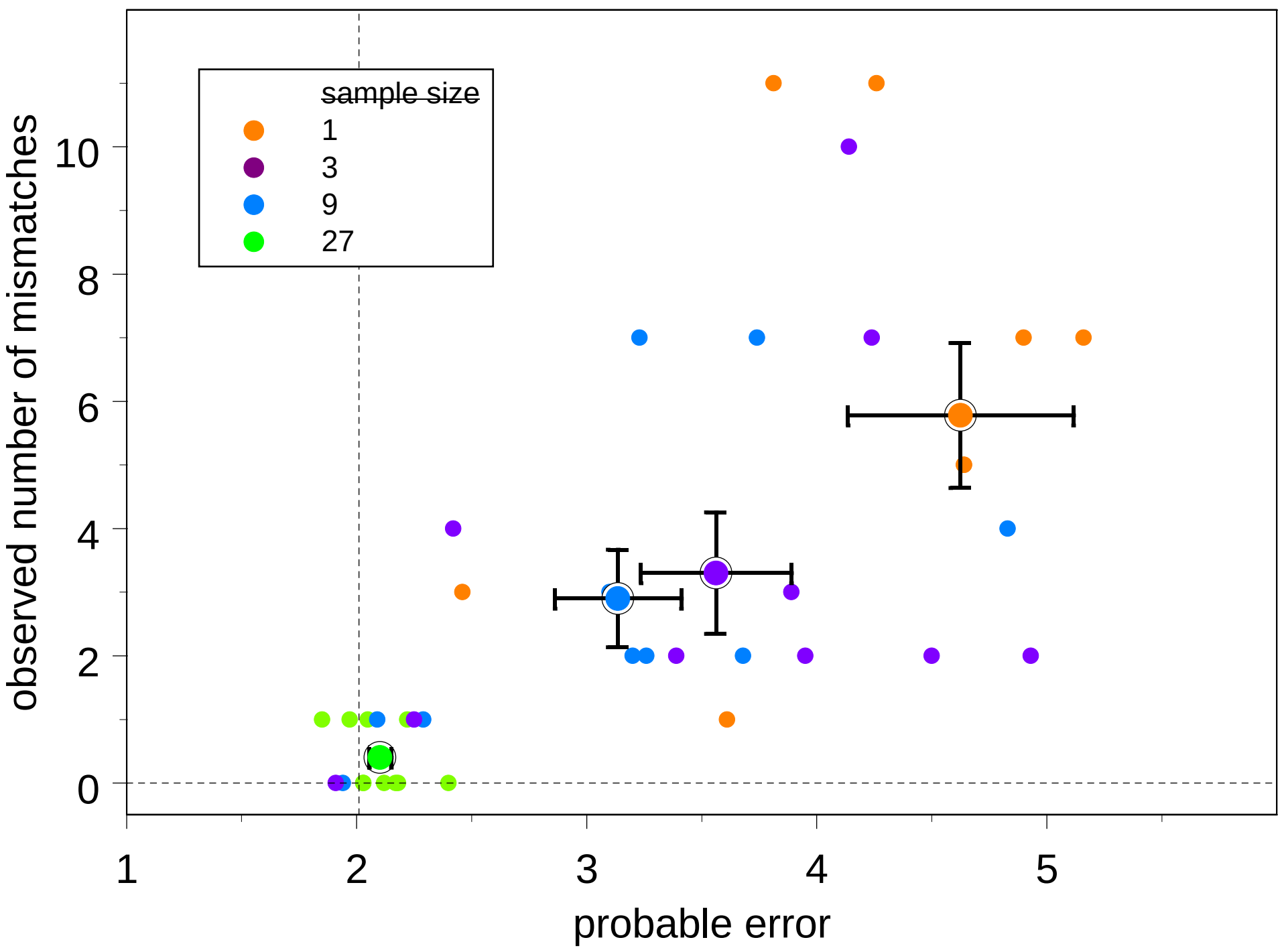

**Figure 5. Observed number of mismatches vs probable error.**
The number of mismatches between the modal UA for each subsample compared to the UA for the Clone H complete set vs the estimated error summed over all positions for each Clone H subsample UA. Symbol colors indicate subsample size as shown in the legend. The larger symbols indicate the means; the half-widths of the error bars are the standard errors of the means. The dashed vertical line indicates the total probable error using the complete 84-sequence set.